\documentclass[conference]{IEEEtran}
\IEEEoverridecommandlockouts
\usepackage{cite}

\ifCLASSINFOpdf
\usepackage[pdftex]{graphicx}
\graphicspath{{./images/}}
\DeclareGraphicsExtensions{.pdf,.jpeg,.png}
\else
\usepackage[dvips]{graphicx}
\graphicspath{{./images/}}
\DeclareGraphicsExtensions{.eps}
\fi

\usepackage{amsmath,amssymb}
\usepackage{algorithmicx}
\usepackage{algorithm,algpseudocode}
\usepackage{textcomp}
\usepackage{xcolor}

\DeclareMathOperator*{\argmax}{argmax}
\newcommand\rv[1]{{\color{black}#1}}

\begin{document}
\title{Centralized Scheduling Strategies for Cooperative HARQ Retransmissions in Multi-Source Multi-Relay Wireless Networks
\thanks{Part of this work has been performed in the framework of the Horizon 2020 project ONE5G (ICT-760809) receiving funds from the European Union. The authors would like to acknowledge the contributions of their colleagues in the project, although the views expressed in this contribution are those of the authors and do not necessarily represent the project.}}
\author{\IEEEauthorblockN{Stefan Cerovi{\'c}\IEEEauthorrefmark{1}\IEEEauthorrefmark{3},
Rapha{\"e}l Visoz\IEEEauthorrefmark{1},
Louis Madier\IEEEauthorrefmark{2} and
Antoine O. Berthet\IEEEauthorrefmark{3}}
\IEEEauthorblockA{\IEEEauthorrefmark{1}Orange Labs, Ch{\^a}tillon, France\\
Email: \{stefan.cerovic, raphael.visoz\}@orange.com}
\IEEEauthorblockA{\IEEEauthorrefmark{2}Nokia, Nozay, France\\
Email: louis.madier@nokia.com}
\IEEEauthorblockA{\IEEEauthorrefmark{3}Department of Telecommunications\\
CENTRALESUPELEC, Gif-sur-Yvette, France\\
Email: antoine.berthet@centralesupelec.fr}}

\IEEEoverridecommandlockouts
\IEEEpubid{\makebox[\columnwidth]{978-1-5386-5541-2/18/\$31.00~\copyright2018 IEEE \hfill} \hspace{\columnsep}\makebox[\columnwidth]{ }}
\maketitle
\IEEEpubidadjcol

\begin{abstract}
In this paper, we investigate centralized scheduling strategies for cooperative incremental redundancy retransmissions in the slow-fading half-duplex multiple access multiple relay channel. Time Division Multiple Access is assumed for the sources and the relays. Sources transmit successively in time slots for the first phase. The second phase consists of a limited number of time slots for retransmissions. In each time slot, the destination schedules a node (being a relay or a source) to retransmit, conditional on the knowledge of the correctly decoded source sets of each node (which is itself for a source). A scheduled relay retransmission uses Joint Network and Channel Coding on its correctly decoded source messages (cooperative retransmission). Several node selection strategies are proposed based on the maximization of the long-term aggregate throughput under a given constraint of fairness. Monte-Carlo simulations show that these strategies outperform the state of the art one based on the minimization of the probability of the common outage event after each time-slot. Moreover, the long-term aggregate throughput reached with these strategies is close to the upper-bound, calculated by the exhaustive search approach. The same conclusion remains valid for both symmetric and asymmetric source rate scenarios.
\end{abstract}

\IEEEpeerreviewmaketitle

\section{Introduction}
The usage of relays in wireless networks allows either to: (1) Extend the coverage of the network; (2) Increase the spectral efficiency of the network (for a fixed power), or transmit with less power (for a fixed throughput). Fundamental principles of cooperative communications can be found in \cite{b1}, where the key idea is that the relay can use the overheard transmission from the source to form its own transmission which helps the decoding at the destination.

In this work, Multiple Access Multiple Relay Channel (MAMRC), denoted by ($M$,$L$,$1$)-MAMRC, consisting of $M\geq 2$ independent users (sources) and L dedicated relays (nodes which do not have their own messages to transmit), is considered. Multiple Access is orthogonal in time (OMAMRC), where transmissions occur in consecutive time-slots. All relays are half-duplex, and apply Selective Decode-and-Forward (SDF) protocol. In SDF, each relay tries to decode the messages of the sources, and sends a function of the correctly decoded ones (usually the error detection is based on CRC added to the source payload before encoding). There, unlike in the classic DF, relays do not have to wait to successfully decode all the sources, as they can cooperate with their correctly decoded subset at a given time. All links are subject to Additive White Gaussian Noise (AWGN) and slow-fading. The existence of limited feedback control channels from the destination to the sources and relays is assumed. That enables the destination to efficiently control the available channel resources. We also assume forward coordination control channels, i.e., each relay can inform the destination about its set of correctly decoded source messages. Incremental Redundancy (IR) Hybrid Automatic Repeat Request (HARQ) is used as an efficient a posteriori fast link adaptation mechanism, i.e., adapting the sources' coding rates to the experienced channel quality. In each time-slot the destination schedules one node (a source or a relay) to transmit. If the selected node is a relay, it performs a cooperative strategy based on the Joint Network Channel Coding/Decoding (JNCC/JNCD) framework \cite{b2}. If it is a source, it performs an incremental redundancy retransmission. Search for an optimal scheduling strategy for a defined performance criterion is the main goal of this paper.  

User scheduling for cooperative multi-source multi-relay networks is considered in \cite{b3}, but only for perfect source-to-relay links where classical DF protocol is used. Performance analysis of cooperative single relay network that uses HARQ mechanism is done in \cite{b4}. IR-HARQ protocol in combination with multi-source multi-relay networks is considered in \cite{b5}, where OMAMRC based on Separate Network Channel Coding/Decoding (SNCC/SNCD) is proposed, which is suboptimal comparing to JNCC/JNCD framework. In \cite{b6}, an information outage analysis of SDF with JNCC/JNCD for the slow-fading ($M$,$L$,$1$)-MAMRC with control channels used by IR-HARQ protocol is conducted. The scheduling (node selection) strategies proposed in the latter work, applicable only for the symmetric rate scenario, are far from being optimal in terms of long-term aggregate throughput, even tough the number of retransmissions is minimized. \rv{In this paper, we investigate both symmetric and asymmetric source rate scenarios. We show that the approach based on the maximization of the number of decoded sources at the destination after each time-slot is beneficial for maximizing the long-term aggregate throughput.}

The remainder of the paper is organized as follows. In section \ref{sec:sys_model}, we describe the system model. In section \ref{sec:prob_form}, long-term aggregate throughput and different outage events involved in its calculation are defined. Three different cooperative HARQ retransmission strategies are proposed in section \ref{sec:strategies}. Numerical results are presented in section \ref{sec:numerical_results}. Finally, in section \ref{sec:conclusion}, we conclude the paper.

\section{System Model}\label{sec:sys_model}
In this paper, slow-fading OMAMRC is considered. Each source $s$, belonging to the set $\mathcal{S}=\{s_1,\dots,s_M\}$, sends its message of $K_s$ information bits ($\textbf{u}_s \in \mathbb{F}_2^{K_s}$) toward the destination (the $M$ sources are mutually independent). \rv{The source messages have equal priority}. $L$ half-duplex dedicated relays, belonging to the set $\mathcal{R}=\{r_1,\dots,r_L\}$, help the destination decoding of source messages (Fig. 1). Time-slotted communication is assumed, where the maximum number of time slots in one frame is set to $M+T_{max}$, where $T_{max}\geq L$ is a system design parameter (Fig. 2). During the first phase consisting of the first $M$ time-slots of the frame, each source transmits in turns its message. The time resource is equally shared between the sources defining a time slot made of 
$N_1$ channel uses. In that phase, the relays listen and try to decode the message of each source relying on a Cyclic Redundancy Check (CRC) code for error detection. 
\begin{figure}[!t]
\centering
\includegraphics[scale=0.3]{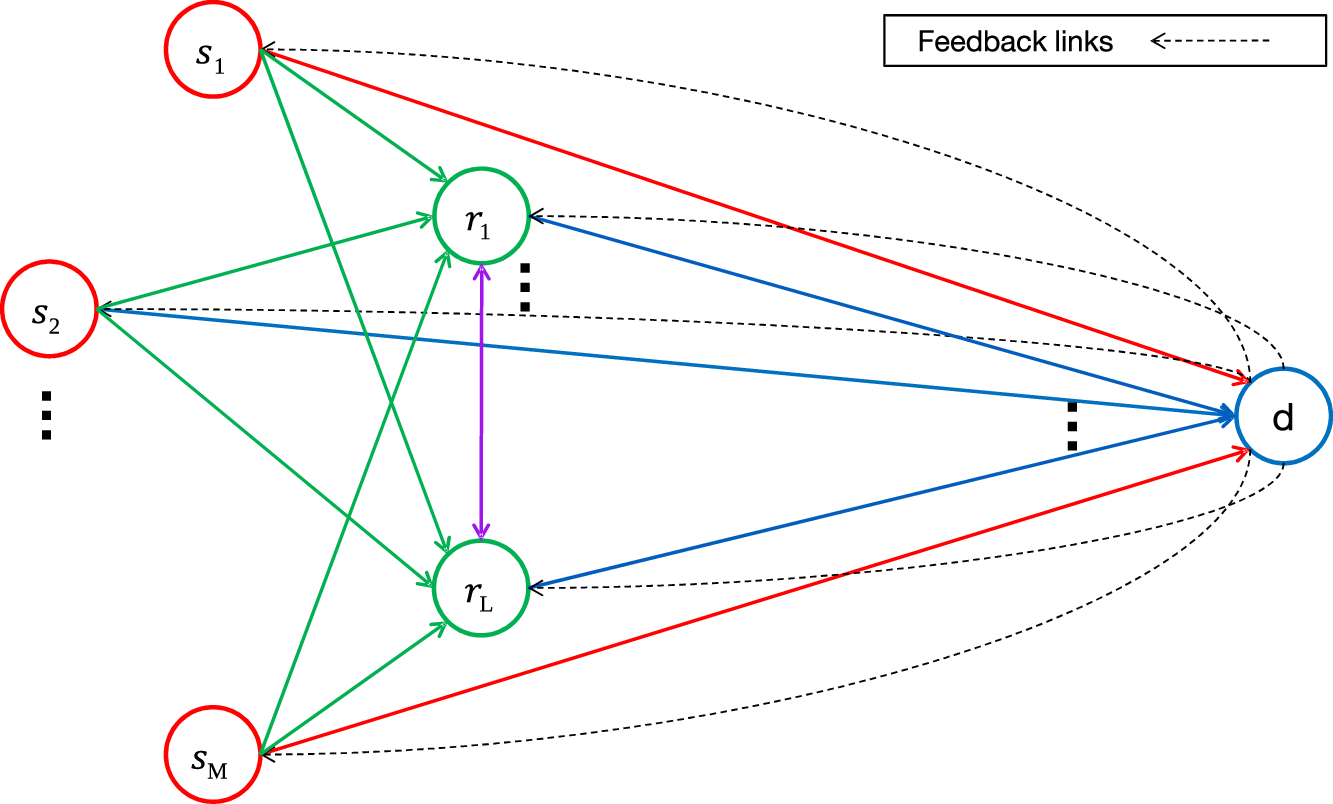}
\caption{Orthogonal Multiple Access Multiple Relay Channel (OMAMRC) with feedback.}
\label{fig:fig_1}
\end{figure}
During the second phase, at the beginning of each time-slot, each relay sends the information about the set of successfully decoded sources to the destination, called the "decoding set" in the following. Forward coordination control channels are used for that purpose, which are assumed to be errorless. Based on that information, in a given time-slot, the destination schedules one node to transmit. The scheduling decision is transmitted to all nodes using errorless limited feedback broadcast control channels. Non-selected relays in a given time-slot can benefit from the transmission of the selected relay. In that way, after each time-slot, the decoding set of each relay (and the destination) can be updated. The exact number $t\in \{1,\dots,T_{max}\}$ of time-slots used in that phase, called also "rounds" in the following, is a random variable and depends on the success of the decoding process at the destination. During the second phase, each time-slot consists of $N_2$ channel uses.
\begin{figure}[!t]
\centering
\includegraphics[scale=0.35]{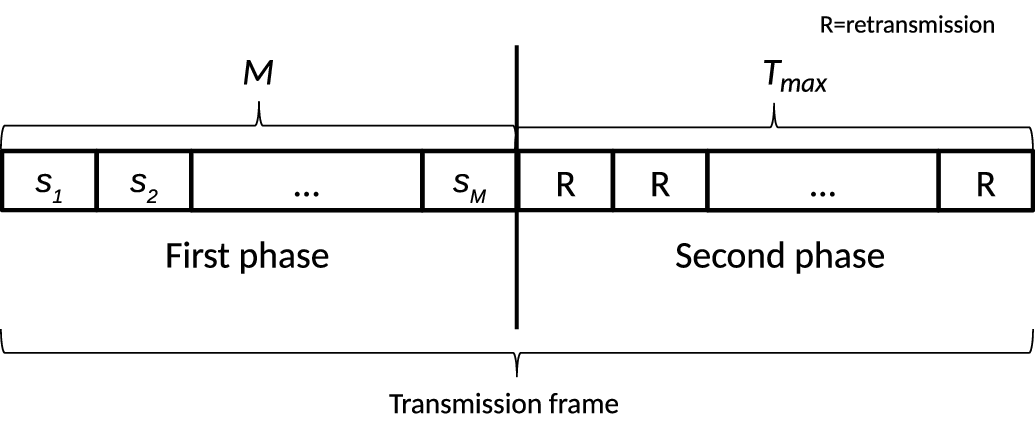}
\caption{One frame is divided into $M+T_{max}$ time-slots.}
\label{fig:fig_2}
\end{figure}
The rates of the sources are fixed during the whole transmission. All nodes transmit with the same power. The following notation is used:
\begin{itemize}
\item $x_{a,k}\in \mathbb{C}$ is the coded modulated symbol for channel use $k$, sent from the node $a\in \mathcal{S}\cup \mathcal{R}$.
\item $y_{a,b,k}$ is a received signal at the node $b\in \mathcal{R}\cup \{d\}$, originating from the node $a$ described previously.
\item $\gamma_{a,b}$ is the average signal-to-noise ratio (SNR) that captures both path-loss and shadowing effects.
\item $h_{a,b}$ are the channel fading gains, which are independent and follow a zero-mean circularly symmetric complex Gaussian distribution with variance $\gamma_{a,b}$.  
\item $n_{a,b,k}$ are independent and identically distributed AWGN samples, which follow a zero-mean circularly-symmetric complex Gaussian distribution with unit variance.
\end{itemize}

Using the previous notation, the received signal at node $b$, originating from node $a$ can be represented as:
\begin{equation}
y_{a,b,k}=h_{a,b}x_{a,k}+n_{a,b,k}.
\end{equation}

During the first phase, $a\in \mathcal{S}$,  $b\in \mathcal{R}\cup \{d\}$, and $k \in \{1,\dots,N_1\}$. During the second phase, $a\in \mathcal{S} \cup \mathcal{R}$, $b\in \mathcal{R}\cup\{d\} \setminus \{a\}$, and $k \in \{1,\dots,N_2\}$.

Only the CSI $\textbf{h}_{\textrm{dir}}=[ h_{s_1,d},\dots,h_{s_M,d},h_{r_1,d},\dots,h_{r_L,d} ] $ of source-to-destination (S-D), and relay-to-destination (R-D) links are perfectly known by the destination. On the other hand, the CSI of source-to-relay (S-R) and relay-to-relay (R-R) links are unknown to the destination. Each node is equipped with one antenna only.

\section{Problem Formulation}\label{sec:prob_form}
Since the destination does not have the CSI of all links in the network, it cannot take an optimal scheduling decision, no matter the criterion of optimization. Let us denote by $\mathcal{S}_{b,t}\subseteq \mathcal{S}$ the decoding set of node $b$ after the round $t\in \{0,\dots,T_{max}\}$. Each relay $r$ at the beginning of the round $t$ sends $\mathcal{S}_{r,t}$ to the destination, which can be considered as a partial knowledge of the CSI of S-R and R-R links. Let $\mathcal{P}_{t-1}$ denote the set collecting the nodes $\hat{a}_l$ which were selected in rounds $l \in \{ 1,\dots,t-1 \}$ prior to round $t$ together with their associated decoding sets $\mathcal{S}_{\hat{a}_l,l-1}$, and the decoding set of the destination $\mathcal{S}_{d,t-1}$. We can consider the end of the first phase as the end of the round zero, where $\mathcal{P}_0$ gathers only sets $\mathcal{S}_{b,0}$, $\forall b\in\ \mathcal{R}\cup \{d\}$. 

Let us define the set comprised of all sources and relays in the network with $\mathcal{N}=\mathcal{S} \cup \mathcal{R}$. For our further analysis it is useful to define the event $\mathcal{E}_t(a_t,\mathcal{S}_{a_t,t-1} |\textbf{h}_{\textrm{dir}},\mathcal{P}_{t-1})$, where at least one source is not decoded correctly at the destination at the end of the round $t$. This event obviously depends on the selected node $a_t\in \mathcal{N}$, and associated decoding set $\mathcal{S}_{a_t,t-1}$ (note that the decoding set of a source is itself). It is conditional on the knowledge of $\textbf{h}_{\textrm{dir}}$ and $\mathcal{P}_{t-1}$. We call this event "the common outage event after round $t$". The individual outage event $\mathcal{O}_{s,t}(a_t,\mathcal{S}_{a_t,t-1} |\textbf{h}_{\textrm{dir}},\mathcal{P}_{t-1})$ is the event that source $s$ is not decoded correctly at the destination after round $t$. The probability of the common outage event $\mathcal{E}_t (a_t,\mathcal{S}_{a_t,t-1} |\textbf{h}_{\textrm{dir}},\mathcal{P}_{t-1})$ for a candidate node $a_t$ can be formulated as $\mathbb{E}(\textbf{1}_{\{\mathcal{E}_t (a_t,\mathcal{S}_{a_t,t-1} |\textbf{h}_{\textrm{dir}},\mathcal{P}_{t-1})\} } )$, where $\mathbb{E}(.)$ is the expectation operator, and $\textbf{1}_{\{\mathcal{V}\}}$ has the value $1$ if the event $\mathcal{V}$ is true, and $0$ otherwise. In the same way the probability of the individual outage event can be defined. In the rest of the paper, in order to simplify the notation, the dependency on $\textbf{h}_{\textrm{dir}}$ and $\mathcal{P}_{t-1}$ is omitted.

Let us define the initial transmission rate of each source $s$ as $R_s=K_s/N_1$ in bit per complex dimension or [b.c.u]. Long-term transmission rate $\bar{R}_s$ per source is defined as the number of transmitted bits over the total number of channel uses spent, for a number of frames that tends to infinity. By averaging out the number of rounds used in the second phase $\mathbb{E}(T)=\sum_{t=1}^{T_{max}}t\textrm{Pr}\{T=t\}$, and by defining $\alpha=N_2 / N_1$, it can be expressed as follows:
\begin{equation}
\bar{R}_s=\frac{R_s}{M+\alpha\mathbb{E}(T)},
\label{eq:first}
\end{equation}

The long-term aggregate throughput can be defined as the sum of the individual throughputs:
\begin{equation}
\eta=\sum_{s \in \mathcal{S}}\bar{R}_{s}(1-\textrm{Pr}\{\mathcal{O}_{s,T_{max}}\}).
\label{eq:third}
\end{equation}

The goal of this paper is to maximize the long-term aggregate throughput by applying the proper centralized scheduling strategy of the sources. \rv{It is understood that the node selection within our centralized scheduling strategies should be fair in the sense that the node selection should not depend on the initial rates of the sources. As a result, the maximization of the long-term aggregate throughput under this fairness constraint is equivalent to the maximization of the normalized long-term aggregate throughput defined as:}
\begin{equation}
\bar{\eta}=\sum_{s \in \mathcal{S}}\frac{1}{M+\alpha\mathbb{E}(T)}(1-\textrm{Pr}\{\mathcal{O}_{s,T_{max}}\}).
\label{eq:fourth}
\end{equation}

\rv{Note that it is equivalent to the maximization of the long-term aggregate throughput in case of symmetric initial rates, i.e., $\bar{R}_{s} = R$ for all $s\in\mathcal{S}$.}

Joint Network Channel Coding/Decoding framework is used. Therefore, in each round $l$ in the second phase, the transmitted sequence of the selected node $\hat{a}_l$ (if it is a relay), and the transmitted sequences of the sources in $\mathcal{S}_{\hat{a}_l,l-1}$ form a joint codeword on the messages of the sources in $\mathcal{S}_{\hat{a}_l,l-1}$.

Let $\bar{\mathcal{S}}_{d,t-1}=\mathcal{S} \setminus {\mathcal{S}}_{d,t-1}$ be the set of non-successfully decoded sources at the destination after the round $t-1$. For any subset $\mathcal{B}\subseteq \bar{\mathcal{S}}_{d,t-1}$ and a given candidate node $a_t$, we define the common outage event of sources in $\mathcal{B}$ after round $t$ if the vector of rates $(R_1,R_2,\dots,R_M)$ lies outside of the corresponding $|\mathcal{B}|$-user MAC capacity region. In that case, the sources that belong to $\mathcal{I}=\bar{\mathcal{S}}_{d,t-1}\setminus \mathcal{B}$ are considered as interference. Analytically, that event can be expressed as:
\begin{equation}
\mathcal{E}_{t,\mathcal{B}}(a_t,\mathcal{S}_{a_t,t-1})=\bigcup_{\mathcal{U}\subseteq \mathcal{B}} \mathcal{F}_{d,\mathcal{B}}(\mathcal{U}),
\label{eq:reduced_mac1}
\end{equation}
where $\mathcal{F}_{d,\mathcal{B}}(\mathcal{U})$ stands for the non-respect of the MAC inequality associated to the sum rates of the sources contained in $\mathcal{U}$:
\begin{equation}
\begin{split}
\mathcal{F}_{d,\mathcal{B}}(\mathcal{U})=&\Big\{ \sum_{s \in \mathcal{U}}R_s > \sum_{s \in \mathcal{U}} I_{s,d}\\
&+ \sum_{l=1}^{t-1} \alpha I_{\hat{a}_l,d} \textbf{1}_{\{\mathcal{C}_{\hat{a}_l,s} \}} + \alpha I_{a_t,d} \textbf{1}_{\{ \mathcal{C}_{a_t,s} \}} \Big\},
\end{split}
\label{eq:reduced_mac2}
\end{equation}
where $\mathcal{C}_{\hat{a}_l,s}=\Big\{ \{s\in \mathcal{S}_{\hat{a}_l,l-1}\cap \mathcal{U}\}\wedge \{\mathcal{S}_{\hat{a}_l,l-1}\cap \mathcal{I}=\emptyset\} \Big\}$, and $\mathcal{C}_{a_t,s}=\Big\{ \{s\in \mathcal{S}_{a_t,t-1}\cap \mathcal{U}\}\wedge \{\mathcal{S}_{a_t,t-1}\cap \mathcal{I}=\emptyset\} \Big\}$ with $\wedge$ standing for the logical and. In (\ref{eq:reduced_mac2}), $I_{a,b}$ represents the mutual information between the nodes $a$ and $b$. Multiplication by $\alpha$ serves as a normalization before adding two mutual information originating from two different phases, where the transmission uses $N_1$ and $N_2$ channel uses, respectively. 

The individual outage event of the source $s$ after the round $t$ can be defined as:
\begin{equation}
\begin{split}
&\mathcal{O}_{s,t}(a_t,\mathcal{S}_{a_t,t-1})=\bigcap_{\mathcal{I}\subset \bar{\mathcal{S}}_{d,t-1}} \bigcup_{\mathcal{U}\subseteq \bar{\mathcal{I}}:s\in \mathcal{U}}\Big\{ \sum_{s \in \mathcal{U}}R_s > \sum_{s \in \mathcal{U}} I_{s,d}\\
 &+ \sum_{l=1}^{t-1} \alpha I_{\hat{a}_l,d} \textbf{1}_{\{\mathcal{C}_{\hat{a}_l,s}\}} + \alpha I_{a_t,d} \textbf{1}_{\{\mathcal{C}_{a_t,s}\}} \Big\},
\label{eq:Ost_JNCC}
\end{split}
\end{equation}
where $\bar{\mathcal{I}}=\bar{\mathcal{S}}_{d,t-1}\setminus \mathcal{I}$, and $\mathcal{C}_{\hat{a}_l,s}$ and $\mathcal{C}_{a_t,s}$ have the same definition as above.

\rv{
\section{Cooperative HARQ retransmission strategies}\label{sec:strategies}
As a summary, in the given round $t$ in the second phase, the following HARQ mechanism occurs:
\begin{enumerate}
\item The destination broadcasts a common ACK bit using a feedback control channel to all the other nodes if it succeeded in decoding all the source messages after round $t-1$. Otherwise it broadcasts a common NACK.
\item If a NACK bit was sent by the destination, each relay $r$ sends its decoding set $\mathcal{S}_{r,t-1}$ using forward coordination channels. Otherwise, if an ACK bit was sent, a new frame begins and the sources transmit new messages while the relays and destination empty their memory.
\item Based on a partial knowledge of the channel matrix, $\textbf{h}_{\textrm{dir}}$, and the set $\mathcal{P}_{t-1}$, the destination makes a scheduling decision about the node to select for transmission. Its decision is broadcasted using a feedback control channel.
\item Selected node transmits. If a source is selected, it sends additional redundancy bits of its original message. If a relay is selected, it performs JNCC with the messages of the sources that it was able to decode.
\end{enumerate}}

The practical approach of the selection strategies described in \cite{b6}, which are based on the minimization of the probability of the common outage event after each round in the second phase, seems as a good idea at first sight for the maximization of the long-term aggregate throughput. Indeed, since $\textrm{Pr}\{\mathcal{O}_{s,T_{max}}\}\leq \textrm{Pr}\{\mathcal{E}_{T_{max}}\}$ for each $s\in\mathcal{S}$, and since $\textrm{Pr}\{\mathcal{E}_t\}\leq \textrm{Pr}\{\mathcal{E}_{t-1}\}$, the individual outage probabilities of all sources are being lowered as well, while the average number of retransmission rounds in the second phase is minimized. Note that $\textrm{Pr}\{\mathcal{O}_{s,T_{max}}\}$ and $\mathbb{E}(T)$ are in the expression of the long-term aggregate throughput. However, the individual outage probabilities are not minimized in that way (even tough they are lowered), which can be very costly in the case where the quality of each link in the network is bad. In such a scenario, we can end up easily in the case where neither of the sources is being decoded correctly at the end of the round $T_{max}$. \rv{Hence, it may be more useful to dedicate the retransmission rounds to the successful decoding of a subset of sources. As shown in Section \ref{sec:numerical_results}, it has a positive impact on the long-term aggregate throughput. Another drawback of the proposed strategies in \cite{b6} is that they are applicable only to the symmetric source rate scenario.}

\rv{In the following, we propose three different node selection strategies.}

\subsection{Strategy 1: Node selection based on the number of newly decoded sources}
The idea of this strategy is to go exhaustively through all node selection alternatives and try to find the node activation that maximizes the number of decoded sources at the destination at the end of round $t$. The derivation of the number of decoded sources is performed only for the nodes for which $\bar{\mathcal{S}}_{d,t-1}\cap \mathcal{S}_{a_t,t-1}\neq \emptyset$. The destination chooses the node which brings the highest number of newly decoded sources, which is equivalent to the maximization of the cardinality of the decoding set of the destination. When there are multiple nodes that bring the same number of decoded sources at the destination, the selected node is the one having the highest mutual information between itself and the destination. 

The reasoning for that comes from the nature of (\ref{eq:Ost_JNCC}), where if we want to minimize the probability of the individual outage event of a given source $s$, we need to maximize the right-hand side of all the inequalities that are in (\ref{eq:Ost_JNCC}). Obviously, the choice of the candidate node $a_t$ only affects $\alpha I_{a_t,d} \textbf{1}_{\{\mathcal{C}_{a_t,s}\}}$ in (\ref{eq:Ost_JNCC}). The problem is that in general, there is not a single node $a_t$ that simultaneously maximizes $\alpha I_{a_t,d} \textbf{1}_{\{\mathcal{C}_{a_t,s}\}}$ for each possible $s$, $\mathcal{I}$ and $\mathcal{U}$. However, for a given $s$, $\mathcal{I}$ and $\mathcal{U}$, the optimal selection process is equivalent to the choice of the candidate node $a_t$ with the highest $I_{a_t,d}$, for which $\mathcal{C}_{a_t,s}=1$. Therefore, our intuitive approach for the node selection is:
\begin{equation}
\hat{a}_t=\argmax_{a_t\in\mathcal{A}^\prime_t}\{ I_{a_t,d}\},
\end{equation}
where $\mathcal{A}^\prime_t$ is the set of candidate nodes that maximizes the decoding set of the destination after the round $t$.

Our method for determining the decoding set of the destination for the selected candidate node $a_t$ is based on multiple checks of common outage events associated to the subsets of $\bar{\mathcal{S}}_{d,t-1}$. \rv{Before proceeding to the algorithm, let us recall the sufficient and necessary conditions for a set of sources to be the decoding set of the destination. They are: (i) The $|\mathcal{S}_d|$-user MAC is not in common outage and (ii) For all the subsets $\mathcal{S}_d^\prime$ that include $\mathcal{S}_d$ ($\mathcal{S}_d\subset \mathcal{S}_d^\prime$), the $|\mathcal{S}_d^\prime|$-user MAC is in outage. }

Let us denote with $\mathcal{B}_j^{(i)}$ the $j$-th subset of cardinality $i$ of the $\bar{\mathcal{S}}_{d,t-1}$, where $j\in \{1,\dots,\binom{|\bar{\mathcal{S}}_{d,t-1}|}{i}\}$  (as there are total of $\binom{|\bar{\mathcal{S}}_{d,t-1}|}{i}$ subsets of cardinality $i$ in the $\bar{\mathcal{S}}_{d,t-1}$). Furthermore, let us denote with $v(a_t)$, where $a_t\in \mathcal{N}$, the number of newly decoded sources at the destination after round $t$ comparing to round $t-1$, by choosing the node $a_t$. The step-by-step algorithm of this strategy is described in Alg. \ref{alg:Strategy_1}. 
\begin{algorithm}[thb]
\caption{Node selection process of strategy $1$.}\label{alg:Strategy_1}
\begin{algorithmic}[1]
\State $v(a_t^\prime)\gets 0$, $\forall a_t^\prime\in \mathcal{N}$.\Comment{Initialization.}
\State $\textrm{max}_v\gets 0$.\Comment{We track the maximum $v(a_t)$.}
\For{$n\gets 1$ \textbf{to} $|\mathcal{N}|$}
	\State $a_t\gets \mathcal{N}(n)$.\Comment{Pick a new candidate node.}
	\If{$\bar{\mathcal{S}}_{d,t-1}\cap \mathcal{S}_{a_t,t-1}=\emptyset$}
	\State \textbf{continue}.\Comment{$v(a_t)=0$ remains.}
	\EndIf
	\State $\textrm{found}\gets 0$.\Comment{Indicator that we found $v(a_t)$.}
	\For {$i\gets |\bar{\mathcal{S}}_{d,t-1}|$ \textbf{to} $1$}
		\For {$j\gets 1$ \textbf{to} $\binom{|\bar{\mathcal{S}}_{d,t-1}|}{i}$}
			\State Calculate $\mathcal{E}_{t,\mathcal{B}_j^{(i)}}(a_t,\mathcal{S}_{a_t,t-1})$ (using (\ref{eq:reduced_mac1}), (\ref{eq:reduced_mac2})).
			\If{$\mathcal{E}_{t,\mathcal{B}_j^{(i)}}(a_t,\mathcal{S}_{a_t,t-1})=0$}
				\State $v(a_t)\gets i$.
				\If{$v(a_t)>\textrm{max}_v$}
					\State $\textrm{max}_v=v(a_t).$
				\EndIf
				\State $\textrm{found}\gets 1$.
				\State \textbf{break}.
			\EndIf
		\EndFor
		\If{$\textrm{found}=1$}	
			\State \textbf{break}.
		\EndIf	
	\EndFor
\EndFor
\State $\mathcal{A}^\prime_t \gets \emptyset$.\Comment{Set of candidate nodes with maximum $v(a_t)$.}
\For{$n\gets 1$ \textbf{to} $|\mathcal{N}|$}
	\State $a_t\gets \mathcal{N}(n)$.
	\If{$v(a_t)=\textrm{max}_v$}
		\State $\mathcal{A}^\prime_t\gets \mathcal{A}^\prime_t \cup \{a_t\}$.
	\EndIf
\EndFor
\State $\hat{a}_t\gets \argmax_{a_t\in \mathcal{A}^\prime_t} \{I_{a_t,d}\}$.
\end{algorithmic}
\end{algorithm}

\subsection{Strategy 2: Node selection based on the highest mutual information}
In this strategy, the similar intuitive approach is used as in strategy $1$ for the case where multiple nodes can provide the destination with the same number of newly decoded sources. The difference here is that in round $t$, each node that was able to decode at least one source from the set $\bar{\mathcal{S}}_{d,t-1}$ is a candidate node. That is to say, the selection criterion has the following form:
\begin{equation}
\hat{a}_t=\argmax_{a_t\in\mathcal{S}\cup\mathcal{R}}\{ I_{a_t,d} \textbf{1}_{\{ \bar{\mathcal{S}}_{d,t-1}\cap \mathcal{S}_{a_t,t-1} \neq \emptyset \}}\}.
\label{eq:strategy1}
\end{equation}

\rv{Obviously, this strategy offers much less computational complexity compared with the previous one.}

\subsection{Strategy 3: Node selection based on the highest product of the mutual information and the cardinality of the decoding set}
The biggest drawback of the strategy $2$ is that a node with small cardinality of the set $|\mathcal{S}_{a_t,t-1}|$ may be chosen. So we propose a modification of that strategy where in each round the destination selects the node $a_t$ with the highest product of $I_{a_t,d}\cdot |\mathcal{S}_{a_t,t-1}|$. Such product could potentially be a good joint indicator of both the amount of the mutual information $I_{a_t,d}$, and the cardinality of the decoding set $|\mathcal{S}_{a_t,t-1}|$.

\subsection{The exhaustive search approach for the best possible activation sequence}\label{subs:exhaustive}
Conditional on the knowledge of the CSI of all links in the network, we can find the optimal activation sequence of nodes with respect to normalized long-term aggregate throughput by using the exhaustive search approach. Since the maximum number of rounds is $T_{max}$ and for each round there are $M+L$ possible candidate nodes, the number of possible activation sequences is equal to $(M+L)^{T_{max}}$. For each possible activation sequence, we can check how many sources the destination can decode at any given round. Finally, in order to determine which activation sequence is optimal, the following procedure is used:
\rv{\begin{itemize}
\item Out of all activation sequences leading to the correct decoding of all the sources at the destination, select the one(s) which necessitates the lowest number of rounds. If there are several activation sequence candidates, choose one of them randomly.
\item If no activation sequence leads to the correct decoding of all sources until $T_{max}$, then choose the one which is associated with the highest cardinality of the decoding set at $T_{max}$. If there are still several activation sequence candidates (which brings the same $|\mathcal{S}_{d,T_{max}}|$), choose one of them randomly.
\end{itemize}}

This kind of procedure is computationally very expensive. In addition, we should stress that the knowledge of the CSI of all links would incur extremely large feedback overhead. Thus this strategy has no interest in practice. It is used in Section \ref{sec:numerical_results} as an upper bound yielding the optimal node selection strategy in case of full CSI knowledge.

\section {Numerical Results}\label{sec:numerical_results}
\rv{In this section, we want to evaluate the performance of the three proposed selection strategies in terms of long-term aggregate throughput. Two different selection strategies are used as benchmark. The first one, referred to as "Reference $1$" in the figure legends, is the strategy $1$ from \cite{b6}, which is described in section \ref{sec:strategies}. The second one uses the exhaustive search approach to find the optimal node selection strategy. It is described in section \ref{subs:exhaustive}, and referred to as "Upper-bound" in the figure legends.

We focus on the (3,3,1)-OMAMRC, with $T_{max}=3$ and $\alpha=0.5$. Independent Gaussian distributed channel inputs is assumed (with zero mean and unit variance), in which case $I_{a,b}=\log_2(1+|h_{a,b}|^2)$. Some other formulas could be also used for calculating $I_{a,b}$ taking into account, for example, discrete entries, finite length of the codewords, non-outage achieving JNCC/JNCD architectures etc. They would not have any impact on the basic concepts of this work. 

In the first part of simulations, we assume a symmetric rate scenario where $R_s=R=1$ [b.c.u] for all $s$. All the links in the network are symmetric, i.e., $\gamma_{a,b}=\gamma$, $\forall a\in \mathcal{S} \cup \mathcal{R}$, $\forall b\in \mathcal{R} \cup \{d\}$, and $a \neq b$. Fig. \ref{fig:figs1} shows the long-term aggregate throughput as a function of $\gamma$ for different strategies.
\begin{figure}[!t]
\centering
\includegraphics[scale=0.37]{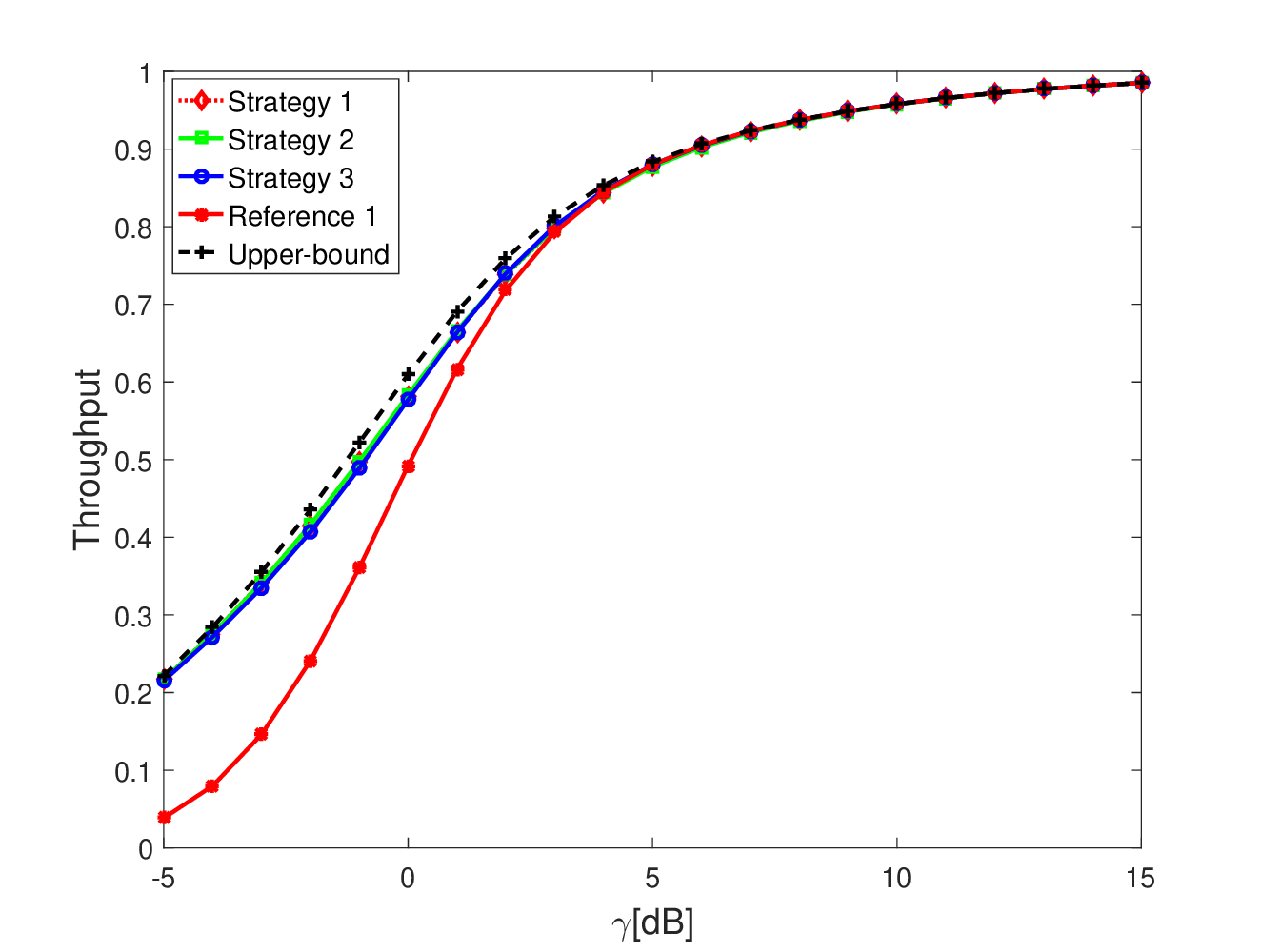}
\caption{Long-term aggregate throughput of different strategies for symmetric rates equal to $R=1[b.c.u]$, and symmetric links scenario.}
\label{fig:figs1}
\end{figure}
In the range of low SNR, "Reference $1$" strategy is significantly worse in terms of long-term aggregate throughput than all the other strategies. Indeed, the minimization of the common outage probability at each round often leads to smaller number of correctly decoded source messages than for the other strategies. Note that the asymptotic limit of the long-term aggregate throughput for boundless capacity links in the network is equal to $\eta=\sum_{s \in \mathcal{S}}R_s/M$, since in that regime $\mathbb{E}(T)\rightarrow 0$ and $\textrm{Pr}\{\mathcal{O}_{s,T_{max}}\}\rightarrow 0$, $\forall s\in \mathcal{S}$. The three strategies proposed in this paper perform close to each other. Strategy $1$ is the best one, and strategy $3$ is the worst. We can conclude that strategies $2$ and $3$ represent a good trade-off between computational complexity and performance. Finally, all the proposed strategies are much closer to the theoretical upper-bound than "Reference $1$" strategy, confirming the validity of our intuitive approach.

The same comparison is made keeping a symmetric link scenario in Fig. \ref{fig:figs2}. However, here the initial rates are chosen from a discrete Modulation and coding scheme (MCS) family whose rates belong to $\{$0.5$,$1$,$1.5$,$2$,$2.5$,$3$,$3.5$\}$ [b.c.u]. The initial rates are chosen to maximize the long-term aggregate throughput with respect to the average SNR (slow link adaptation). Here, the slow link adaptation is very simple since the rates of the sources and the SNR are equal. The slow link adaptation for "Reference $1$" is illustrated by the black dotted line that corresponds each to a given initial rate (the same for each source) ranging from $0.5$ to $3.5$ [b.c.u]. It simply takes the envelope of these curves. The gain of the proposed strategies (within the framework of slow link adaptation) compared with "Reference $1$" is approximately $1$ dB. 
\begin{figure}[!t]
\centering
\includegraphics[scale=0.39]{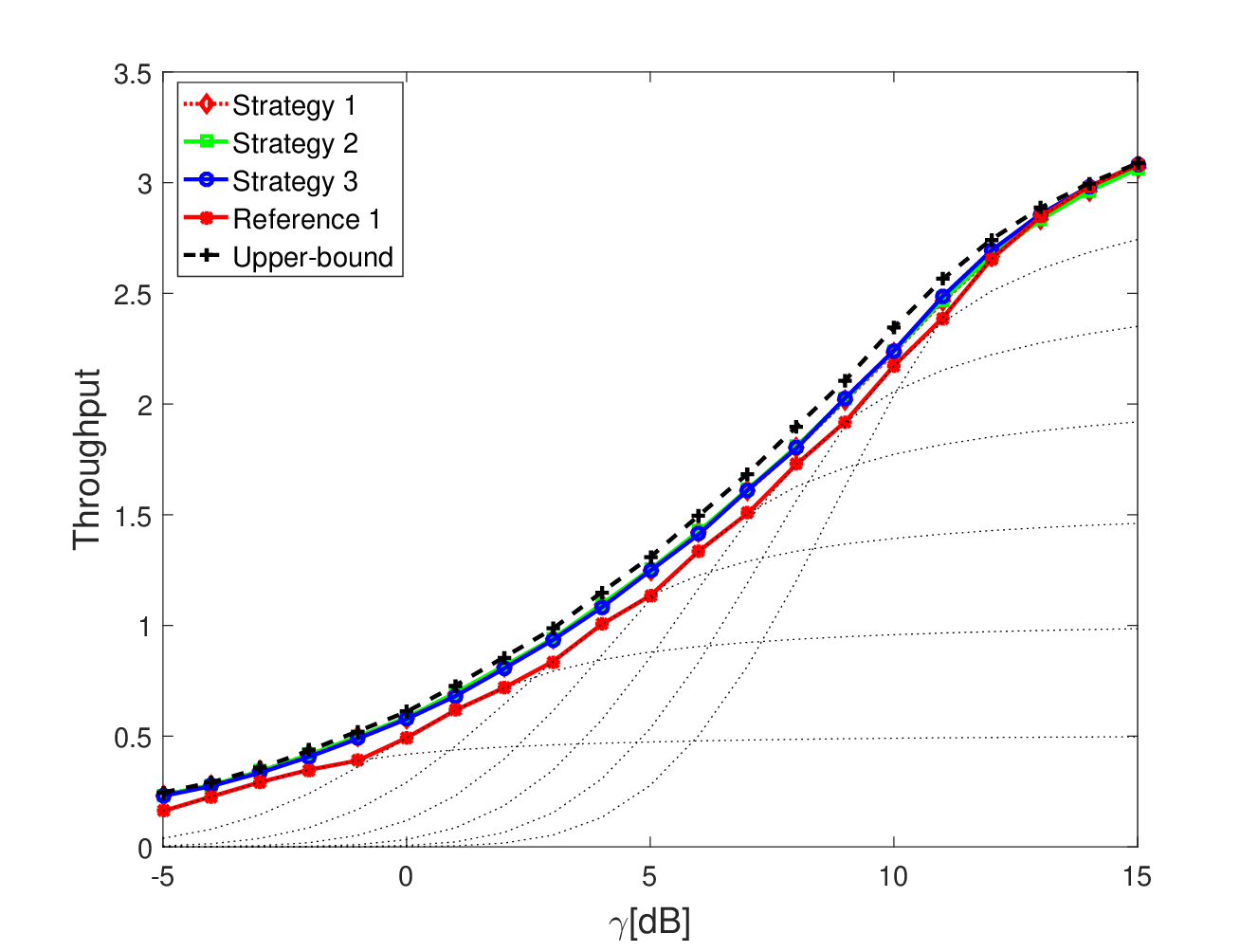}
\caption{Long-term aggregate throughput of different strategies with slow link adaptation and symmetric link scenario.}
\label{fig:figs2}
\end{figure}

Finally, in the second part of the simulations, we want to illustrate the application of the proposed strategies in an asymmetric source rate scenario, where the initial rates are set to $[R_{s_1}=3, R_{s_2}=2.5, R_{s_3}=2]$ [b.c.u] as an example. Average SNR of all links in the network are also set to be asymmetric and in the range: $\gamma_{a,b} \in \{-10\textrm{dB}, \dots, 15\textrm{dB}\}$. This time, the long-term aggregate throughput, shown on the Fig. \ref{fig:figs4}, is a function of $\Delta_{\gamma}$, which is added to each individual link simultaneously with respect to the starting asymmetric link configuration. Obviously, "Reference $1$" strategy is left out from the simulations. We observe that the performance of the proposed strategies remains close to the upper-bound.
\begin{figure}[!t]
\centering
\includegraphics[scale=0.39]{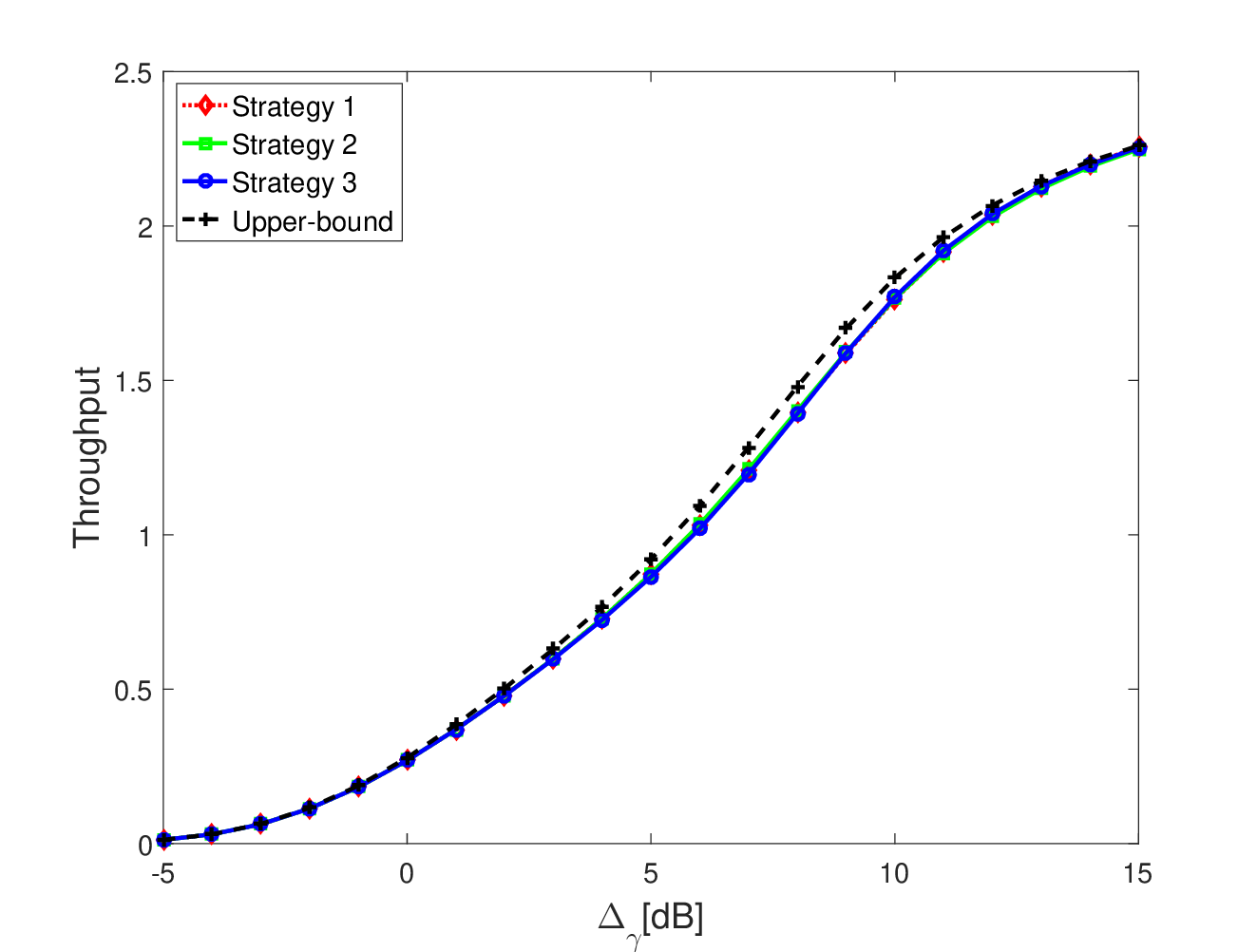}
\caption{Long-term aggregate throughput of different strategies for asymmetric rates $[R_{s_1}=3, R_{s_2}=2.5, R_{s_3}=2]$[b.c.u], and asymmetric link scenario.}
\label{fig:figs4}
\end{figure} }

\section{Conclusion}
\label{sec:conclusion}
\rv{In this paper, we have proposed three node selection strategies for the slow-fading half-duplex time-slotted OMAMRC, that are applicable both to symmetric and asymmetric source rate scenarios. The first one is based on the exhaustive search of the node which maximizes the cardinality of the decoding set at the destination after each time slot. The other two are less computationally expensive and proceed from intuitive approaches. Simulation results show that all the proposed strategies perform close to each other in terms of long-term aggregate throughput. Additionally, our strategies perform close to the long-term aggregate throughput upper bound (under the given fairness constraint) obtained by an exhaustive search over all possible node sequence activations. It confirms the validity of our intuitive approaches. Finally, it is demonstrated that our strategies always perform better than the strategy based on the minimization of the common outage event probability after each time-slot. As a future work, we will investigate efficient slow link adaptation algorithms in the asymmetric link scenario with respect to different quality of services.}


\begin{thebibliography}{1}
\bibitem{b1} T. Cover and A. El Gamal, ``Capacity Theorems for the Relay Channel,'' \textit{IEEE Trans. Inf. Theory}, Vol. IT-25, No. 5, pp. 572-584, Sept. 1979.
\bibitem{b2} C. Hausl and P. Dupraz, "Joint Network-Channel Coding for the Multiple-Access Relay Channel," in \textit{Proc. 3rd Annual IEEE Communications Society on Sensor and Ad Hoc Communications and Networks}, vol. 3, pp. 817-822, Sept. 2006.
\bibitem{b3} Li Wang, X. Zhang and Yuhan Dong, ``User scheduling and relay selection with fairness concerns in multi-source cooperative networks,'' \textit{IEEE WiOpt'13}, Tsukuba Science City, Japan, May 2013.
\bibitem{b4} Y. Ai and M. Cheffena, ``Performance Analysis of Hybrid-ARQ with Chase Combining over Cooperative Relay Network with Asymmetric Fading Channels,'' \textit{IEEE VTCF'16}, Montreal, QC, Sept. 2016.
\bibitem{b5} Y. Cheng and L. Yang, ``Joint relay ordering and linear finite field network coding for multiple-source multiple-relay wireless sensor networks,'' \textit{Int. J. Distrib. Sensor Netw.}, vol. 2015, Sep. 2013. Art. no. 729869.
\bibitem{b6} A. Mohamad, R. Visoz and A. O. Berthet, ``Cooperative Incremental Redundancy Hybrid Automatic Repeat Request Strategies for Multi-Source Multi-Relay Wireless Networks,'' \textit{IEEE Commun. Lett.}, vol. 20, no. 9, pp. 1808-1811, Sept. 2016.
\end{thebibliography}
\end{document}